\documentstyle[11pt]{article}
\def\fnote#1#2{\begingroup\def\thefootnote{#1}\footnote{#2}\addtocounter
{footnote}{-1}\endgroup}

\begin{document}

\hfill{UTTG-04-10}

\vspace{36pt}

\begin{center}
{\large {\bf {Six-dimensional Methods for  Four-dimensional Conformal Field Theories}}}

\vspace{36pt}
Steven Weinberg\fnote{*}{Electronic address:
weinberg@physics.utexas.edu}\\
{\em Theory Group, Department of Physics, University of
Texas\\
Austin, TX, 78712}

\vspace{30pt}

\noindent
{\bf Abstract}
\end{center}
\noindent
The calculation of  both spinor and tensor Green's functions in four-dimensional conformally invariant field theories can be greatly simplified by six-dimensional methods.  For this purpose, four-dimensional fields are constructed as projections of fields on the hypercone in six-dimensional projective space, satisfying certain transversality conditions.  In this way some Green's functions in conformal field theories are shown to have structures more general than those commonly found by use of the inversion operator.  These methods fit in well with the assumption of AdS/CFT duality.  In particular, it is transparent that if fields  on AdS$_5$ approach finite limits on the boundary of AdS$_5$, then in the conformal field theory on this boundary these limits transform with conformal dimensionality zero if they are tensors (of any rank), but with conformal dimension  1/2 if they are spinors or spinor-tensors. 

\vfill

\pagebreak

\begin{center}
{\bf I. INTRODUCTION}
\end{center}

Let's first review some well-known fundamentals.  The action of conformal transformations in four spacetime dimensions on a general field $\psi^n(x)$ is given by its commutators with the generators $J^{\mu\nu}$ of Lorentz transformations, $P^\mu$ of translations, $K^\mu$ of special conformal transformations, and $S$ of dilatations:
\begin{eqnarray}
i\left[J^{\rho\sigma},\psi^n(x)\right]&=&\left(x^\sigma\frac{\partial}{\partial x_\rho}-x^\rho\frac{\partial}{\partial x_\sigma}\right)\psi^n(x) -i\left(j^{\rho\sigma}\right)^n{}_m
\psi^m(x)\;,\\
i\left[P^{\rho},\psi^n(x)\right]&=&-\frac{\partial}{\partial x_\rho}\psi^n(x)\;,\\
i\left[K^{\rho},\psi^n(x)\right]&=&\left(2x^\rho x^\lambda\frac{\partial}{\partial x_\lambda}-x^2
\frac{\partial}{\partial x_\rho}\right)\,\psi^n(x)\nonumber\\&&-2i x_\lambda\left(j^{\lambda\rho}\right)^n{}_m\psi^m(x)+2d\,x^\rho \psi^n(x)\;,\\
i\left[S,\psi^n(x)\right]&=&\left(x^\lambda \frac{\partial}{\partial x^\lambda}+d\right)\,\psi^n(x)\;,
\end{eqnarray}
where $d$ is the conformal dimensionality of the field, and $j^{\rho\sigma}$ is the appropriate matrix representation of the Lie algebra of the Lorentz group, satisfying the commutation relations
\begin{equation}
i\Big[j^{\mu\nu},j^{\rho\sigma}\Big]=\eta^{\nu\rho}j^{\mu\sigma}-\eta^{\mu\rho}j^{\nu\sigma}
-\eta^{\nu\sigma}j^{\mu\rho}+\eta^{\mu\sigma}j^{\nu\rho}\;.
\end{equation}
We can work out the consequences of conformal symmetry for Green's functions of general fields by direct use of these commutation relations, but this is complicated, especially for non-scalar fields, for which $j^{\rho\sigma}\neq 0$, and for three-point and higher Green's functions.  A widely practiced alternative[1] is to make use of invariance under the a single action of the inversion $x^\mu\mapsto -x^\mu/x^2$, but this is also complicated, and not necessarily valid.  The inversion is not an element of the connected part  of the conformal group, but only an outer automorphism, so that it is possible for the commutation relations (1) through (4) to be satisfied without invariance under the inversion.  This makes no difference for two-point functions, or for some more complicated Green's functions involving only scalar fields, but in Section V  we will see examples  of Green's functions for spinor fields that are not invariant under the inversion, even when the commutation relations (1) through (4) are satisfied.  (These comments do not apply if one acts with the inversion an even number of times, but this gets complicated, and it is not what is usually done in deriving the structure of Green's function.)
  Here we are going to offer a different method for the calculation of Green's functions in four-dimensional conformal field theories, based on very elementary calculations in six dimensions.\footnote{This work was done in preparing a course on quantum field theory given in Spring 2010.  Since the original version of this paper was posted on the hep-th archive, I have learned of previous work in which dynamical equations are assumed for fields in six dimensions, and then used to derive physical field equations in four dimensions.  The literature on this goes back to Dirac[2],  where electromagnetic fields and free spinor fields were considered.  Among the first following Dirac to use this approach were Mack and Salam[3].  Other early references are given in a historical review by Kastrup[4].  Extensive work has been done on six dimensional field equations (including constraints on six-dimensional fields found here) corresponding to realistic theories in four dimensions, by Bars[5].  Related work was done by Ferrara, Grillo, and Gatto[6] for the case of symmetric tensors, and extended to superconformal theories by Ferrara[7].  Of course, much work on the AdS/CFT correspondence deals with related problems[8]. 
In contrast to all this previous work, the aim of the present paper is the modest one of using six dimensional field theories to derive only those properties of Green's functions in four dimensions that follow solely from conformal invariance, with no dynamical assumptions.}  Though no dynamical assumptions are made here, to achieve conformal invariance in four dimensions it is found necessary to specify certain relations between the fields in four dimensions and in six dimensions and to impose constraints on the six-dimensional fields, both of which may prove useful in dynamical theories.
	
It is well known that the connected part of the conformal group in four space-time dimensions form the group $SO(4,2)$, which can be realized as linear transformations in a six-dimensional projective space.  This six-dimensional space is a hypercone,
\begin{equation}
\eta_{KL}X^K\,X^L=0\;,
\end{equation}
where $K$, $L$, etc. run over the values $1,2,3,0,5,6$, and  $\eta_{KL}$ is the metric of the six-dimensional space, a diagonal matrix with non-zero elements
\begin{equation}
\eta_{11}=\eta_{22}=\eta_{33}=\eta_{55}=+1\;,~~\eta_{00}=\eta_{66}=-1\;.
\end{equation}
It is a projective space, in the sense that $\lambda \,X^K$ is identified with $X^K$ for any non-zero $\lambda$.  The connection between six and four dimensions is provided by the formula for the spacetime coordinates $x^\mu$, 
\begin{equation}
x^\mu=\frac{X^\mu}{X^5+X^6}\;,
\end{equation}
where as usual $\mu$, $\nu$, etc. run over the values $1, 2, 3, 0$.
  The conformal group consists of transformations
\begin{equation}
X^K\mapsto \Lambda^K{}_L X^L\;,~~~~~~\eta_{KL}\Lambda^K{}_M\Lambda^L{}_N=\eta_{MN}\;,~~{\rm Det}\Lambda=1
\end{equation}
which generate the group of conformal transformations on the $x^\mu$ given by Eq.~(8).  
The generators $J^{KL}=-J^{LK}$ of these transformations satisfy the commutation relations
\begin{equation}
i\Big[J^{KL},J^{MN}\Big]=\eta^{ML}J^{KN}-\eta^{KM}J^{LN}
-\eta^{LN}J^{KM}+\eta^{KN}J^{ML}\;,
\end{equation}
with the generators of translations, special conformal transformations,and dilatations identified as
\begin{equation}
P^\mu=J^{5\mu}+J^{6\mu},~~~K^\mu=-J^{5\mu}+J^{6\mu},~~~S=J^{65}\;.
\end{equation}
The inversion operation $x^\mu\mapsto -x^\mu/x^2$ is simply the reflection that changes the sign of $X^6$, leaving all other $X^K$ unchanged.  It violates the condition ${\rm Det}\Lambda=+1$, and hence belongs to $O(4,2)$ but not to $SO(4,2)$.

Because of the simplicity of the conformal transformation rule (9), it is very easy to work out the consequences of conformal invariance for the Green's functions of fields in the six dimensional projective space.  We can tell by inspection whether a Green's function of fields in six dimensions is $SO(4,2)$-invariant, in much the same way that we can  tell at a glance whether a Green's function in four spacetime dimensions is Lorentz invariant.  The question, then, is how can we convert information about six-dimensional Green's functions into information about the Green's functions of fields in four-dimensional spacetime?  Fields in six dimensions of course have more components than the corresponding fields in four dimensions; for instance, a six-dimensional tensor of rank $r$ has $6^r$ components, rather than the $4^r$ components in four dimensions, and a spinor field in six dimensions has eight rather than four components.  In order to construct suitable four-dimensional fields from fields in six dimensions, we need both to impose constraints on the fields in six dimensions, and write the four-dimensional fields as suitable projections of the six-dimensional fields.

We show how to do this  for tensor fields in Section II.  In Section III we apply these methods to derive the structure of various Green's functions of tensor fields in four-dimensions.    Section IV deals with spinor fields, and in Section V we find some new results for spinor Green's functions.  

Although the methods of this paper described in Sections II through V do not in any way depend on assumptions about holography, they were in fact inspired by AdS/CFT duality[9], especially as explained by Witten[10].  The  six-dimensional methods introduced here are applied to AdS/CFT duality in Section VI, and used to find the conformal dimensionality $d$ of fields in four-dimensional conformal field theories that arise from fields in five-dimensional anti-de Sitter space that approach finite limits on the boundary of the space.  For general tensors, it has the familiar value $d=0$, but for spinor or spinor-tensor fields it is $d=1/2$. 


\begin{center}
{\bf II. TENSOR FIELDS}
\end{center}
A tensor field $T^{K_1 K_2\dots K_r}(X)$ of rank $r$ in six dimensions has the conformal transformation rule
\begin{equation}
T^{K_1 \dots K_r}(X)\mapsto \Lambda_{L_1}{}^{K_1}\cdots \Lambda_{L_r}{}^{K_r}T^{L_1 \dots L_r}(\Lambda X)\;,
\end{equation}
with $\Lambda$ satisfying Eq.~(9).
(Indices $K$, $L$, etc. are lowered and raised with $\eta_{KL}$ and its inverse $\eta^{KL}$.)
For infinitesimal $SO(4,2)$ transformations, this can be expressed as formulas for the commutators of $T^{K_1 K_2\dots K_r}$  with the generators $J^{KL}$ of these transformations:
\begin{eqnarray}
&&i\left[J^{MN},T^{K_1\dots K_r}(X)\right]=\left(X^N\frac{\partial}{\partial X_M}-X^M\frac{\partial}{\partial X_N}\right)T^{K_1\dots K_r}(X)\nonumber\\&&~~~-i\left({\cal J}^{MN}\right)^{K_1\dots K_r}_{L_1\dots L_r}T^{L_1\dots L_r}(X)\;,
\end{eqnarray}
where ${\cal J}^{MN}$ is the tensor representation of the $SO(4,2)$ algebra:
\begin{eqnarray}
&&i\left({\cal J}^{MN}\right)^{K_1\dots K_r}_{L_1\dots L_r}= \left(\eta^{MK_1}\delta^N_{L_1}-\eta^{NK_1}\delta^M_{L_1}\right)\delta^{K_2}_{L_2}\cdots \delta^{K_r}_{L_r}+\dots\nonumber\\&&
~~~+\left(\eta^{MK_r}\delta^N_{L_r}-\eta^{NK_r}\delta^M_{L_r}\right)\delta^{K_1}_{L_1}\cdots \delta^{K_{r-1}}_{L_{r-1}}\;.
\end{eqnarray}
Because we identify $X^K$ with $\lambda X^K$, $T^{K_1\dots K_r}(X)$ must satisfy a scaling relation
\begin{equation}
T^{K_1\dots K_r}(\lambda X)=\lambda^{-d} T^{K_1\dots K_r}(X)\;,
\end{equation}
where for the present $d$ is just some unknown number. For reasons that will become clear, we also require that the hypercone condition (6) must not be affected by any of the differential operators 
$$T^{K_1\dots K_r}(X)\frac{\partial}{\partial X^{K_1}}\;,\cdots\;, ~~~~T^{K_1\dots K_r}(X)\frac{\partial}{\partial X^{K_r}}\;, $$
 so that $T^{K_1\dots K_r}(X)$ must be {\em transverse} on each index
\begin{equation}
X_{K_1}T^{K_1\dots K_r}(X)=0\;,\cdots\;,~~~~~X_{K_r}T^{K_1\dots K_r}(X)=0\;.
\end{equation}

Now, consider the four-dimensional field
\begin{equation}
t^{\mu_1\dots\mu_r}(x)\equiv (X^5+X^6)^d\; e^{\mu_1}_{K_1}(x)\cdots e^{\mu_r}_{K_r}(x)\;T^{K_1\dots K_r}(X)\;,
\end{equation}
with 
\begin{equation}
e^\mu_\nu(x)\equiv \delta^\mu_\nu\;,~~~e^\mu_5(x)\equiv e^\mu_6(x)\equiv -x^\mu\;.
\end{equation}
Because of the scaling condition (15), the field (17) is only a function of the ratios of the $X^K$, so that when we eliminate $X^5-X^6$ by imposing the hypercone condition (6), the field (17) can indeed be regarded as a function only of the spacetime coordinate $x^\mu$ given by Eq.~(8).  

It is straightforward though tedious to use  Eqs.~(6), (8), (11), (13), (15), and (16) to show directly that  the four-dimensional tensor field given by Eqs.~(17) and (18) does satisfy the conformal transformation rules (1) through (4),  with $\left(j^{\rho\sigma}\right)^{\mu_1\dots\mu_r}_{\nu_1\dots\nu_r}$ here given by the tensor representation of the Lorentz group:
\begin{eqnarray}
&&i\left(j^{\rho\sigma}\right)^{\mu_1\dots \mu_r}_{\nu_1\dots \nu_r}= \left(\eta^{\rho \mu_1}\delta^\sigma_{\nu_1}-\eta^{\sigma \mu_1}\delta^\rho_{\nu_1}\right)\delta^{\mu_2}_{\nu_2}\cdots \delta^{\mu_r}_{\nu_r}+\dots\nonumber\\&&
~~~+\left(\eta^{\rho \mu_r}\delta^\sigma_{\nu_r}-\eta^{\sigma \mu_r}\delta^\rho_{\nu_r}\right)\delta^{\mu_1}_{\nu_1}\cdots \delta^{\mu_{r-1}}_{\nu_{r-1}}\;.
\end{eqnarray}
In this paper we will instead show this by a less direct but more illuminating method.  

It is shown in the 
Appendix that the usual conformal transformation rules of tensor fields just amount to the statement that  under general conformal transformations a tensor of rank $r$ and conformal dimensionality $d$ transforms as a tensor density of weight
\begin{equation}
w=-(d+r)/4\;.
\end{equation}
So this is the condition that must be satisfied by the field (17).  To show that this condition is satisfied, we note by differentiating Eq.~(8) that 
$$
\frac{\partial x^\mu(X)}{\partial X^K}=(X^5+X^6)^{-1}e^\mu_K(x)\;,
$$
so that the field (17) can be written
$$ t^{\mu_1\dots\mu_r}(x)\equiv (X^5+X^6)^{d+r}\;\frac{\partial x^{\mu_1}(X)}{\partial X^{K_1}}\cdots \frac{\partial x^{\mu_r}(X)}{\partial X^{K_r}} \;T^{K_1\dots K_r}(X)\;.
$$
Hence, under a coordinate transformation $X\mapsto X'=\Lambda^K{}_LX^L$, we have 
$$ t^{\mu_1\dots\mu_r}(x)\mapsto (X'^5+X'^6)^{d+r}\;\frac{\partial x^{\mu_1}(X')}{\partial X^{K_1}}\cdots \frac{\partial x^{\mu_r}(X')}{\partial X^{K_r}} \;\Lambda_{L_1}{}^{K_1}\Lambda_{L_r}{}^{K_r}T^{L_1\dots L_r}(X')\;.
$$
Now, for any displacement $dX$ on the hypercone (6), we have
$$ \frac{\partial x^\mu(X')}{\partial X'^L}dX'^L= \frac{\partial x^\mu(X')}{\partial X^K}dX^K
= \frac{\partial x^\mu(X')}{\partial X^K}\Lambda_L{}^K dX'^L \;.
$$
But this is only for $dX'^L$ on the hypercone, i. e. for  $X'_L\,dX'^L=0$, so 
$$
\frac{\partial x^\mu(X')}{\partial X'^L}- \frac{\partial x^\mu(X')}{\partial X^K}\Lambda_L{}^K \propto X'_L\;.
$$
Under the transversality condition (16) the term proportional to $X'_L$ makes no contribution, so we see that
$\frac{\partial x^{\mu_1}(X)}{\partial X^{K_1}}\cdots \frac{\partial x^{\mu_r}(X)}{\partial X^{K_r}} \;T^{K_1\dots K_r}(X)$ transforms as a tensor under general conformal transformations.  
Furthermore, it is straightforward to show that under general conformal transformations $x\mapsto x'$, the quantity $X^5+X^6$ transforms as a scalar density of weight $-1/4$:
$$ \frac{X'^5+X'^6}{X^5+X^6}=\left|\frac{\partial x'}{\partial x}\right|^{-1/4}\;. $$
Hence $t^{\mu_1\dots\mu_r}(x)$ does indeed transform under general conformal transformations as a tensor density of weight 
given by Eq.~(20), the condition for  conformal invariance.

It may be noted that $e^\mu_K(x)X^K=0$, so $t^{\mu_1\dots\mu_r}(x)$ is unchanged if we shift $T^{K_1\dots K_r}(X)$ by an amount proportional to any of $X^{K_1}$ or $X^{K_2}$ etc. This lowers the number of physically relevant components of    $T^{K_1\dots K_r}(X)$ from $6^r$ to $5^r$, and the transversality conditions (16) lowers it further to $4^r$, the appropriate number for a four-dimensional tensor of rank $r$.

It may also be noted, as a consequence of Eq.~(16), that traces of the four-dimensional tensor $t^{\mu_1\dots\mu_r}(x)$ are proportional to the corresponding traces of the six-dimensional tensor $T^{K_1\dots K_r}(X)$.  For instance,
$$ \eta_{\mu_1\mu_2}t^{\mu_1\mu_2\dots\mu_r}(x)=(X^5+X^5)^d \,e^{\mu_3}_{K_3}(x)\,e^{\mu_4}_{K_4}(x)\cdots \eta_{K_1K_2}\,T^{K_1 K_2\dots K_r}(X)\;.$$
In particular, the condition of being traceless carries over from a six-dimensional tensor $T^{K_1\dots K_r}(X)$ to the corresponding four-dimensional tensor $t^{\mu_1\dots\mu_r}(x).$  The same is obviously also true for conditions of symmetry or antisymmetry.    Hence 
six-dimensional tensors belonging to irreducible representations of $SO(4,2)$ yield four-dimensional tensors  belonging to the corresponding irreducible representations of $SO(3,1)$.
  
\begin{center}
{\bf III. TENSOR APPLICATIONS}
\end{center}

We will first apply the method described in the previous section to a few familiar simple examples, and then turn to more complicated applications.  

\vspace{10 pt}

\noindent
{\bf A. Scalar Fields}

\vspace{10pt}

First, consider the Green's function $\left<\varphi_1(x)\varphi_2(y)\right>_0$ for a pair of scalar fields $\varphi_1(x)$ and $\varphi_2(y)$ of conformal dimensionality $d_1$ and $d_2$, with $x-y$ spacelike.  According to the scaling condition (15), the Green's function for the corresponding six-dimensional fields $\Phi_1(X)$ and $\Phi_2(Y)$ must be of order $-d_1$ in $X$ and $-d_2$ in $Y$, but it can only depend on the scalar $X\cdot Y$, so there must be an equal number of factors of $X$ and $Y$, and therefore $d_1=d_2\equiv d$.  As is well known, this is the one thing beyond scale invariance that we learn in this case from conformal symmetry.  To check that the Green's function in four dimensions has the familiar form dictated by Poincar\'{e} and scale invariance, we note by using Eq.~(8) that the scalar here is
\begin{eqnarray}
&&X\cdot Y= X_\mu Y^\mu+\frac{1}{2}(X^5+X^6)( Y^5-Y^6)+\frac{1}{2}(X^5-X^6)( Y^5+Y^6)\nonumber\\&&=(X^5+X^6)(Y^5+Y^6)\left(x\cdot y-\frac{x^2}{2}-\frac{y^2}{2}\right)=-\frac{1}{2}(X^5+X^6)(Y^5+Y^6)(x-y)^2\;,\nonumber\\&&{}
\end{eqnarray}
so the six dimensional Green's function is proportional to
$$ (X\cdot Y)^{-d}=\left[-\frac{1}{2}(X^5+X^6)(Y^5+Y^6)(x-y)^2\right]^{-d}\;. $$
But according to Eq.~(17), the four-dimensional scalars are related to the six-dimensional scalars by
\begin{equation}
\varphi_1(x)=(X^5+X^6)^d\Phi_1(X)\;,~~~\varphi_2(y)=(Y^5+Y^6)^d\Phi_2(Y)\;,
\end{equation}
so the factors $X^5+X^6$ and $Y^5+Y^6$ cancel in the four-dimensional Green's function, which we see is proportional to $[(x-y)^2]^{-d}$, the well-known result of Poincar\'{e} and scale invariance.

It is almost as easy to deal with the three-point function  $\left<\varphi_1(x)\varphi_2(y)\varphi_3(z)\right>_0$.  According to the scaling condition (15), the corresponding six-dimensional three-point  function for $\Phi_1(X)$, $\Phi_2(Y)$ , and $\Phi_3(Z)$ must be of order $-d_1$ in $X$, $-d_2$ in $Y$, and $-d_2$ in $Z$, so it must be proportional to 
$$ (X\cdot Y)^{-a}(Y\cdot Z)^{-b}(Z\cdot X)^{-c}\;,$$
where $a+c=d_1$, $a+b=d_2$, and $b+c=d_3$, and thus must be proportional to
\begin{eqnarray*}
 &&(X\cdot Y)^{(d_3-d_1-d_2)/2}(Y\cdot Z)^{(d_1-d_2-d_3)/2}(Z\cdot X)^{(d_2-d_1-d_3)/2}\\&&\propto (X^5+X^6)^{-d_1}(Y^5+Y^6)^{-d_2}(Z^5+Z^6)^{-d_3}\\ && \times ((x-y)^2)^{(d_3-d_1-d_2)/2}((y-z)^2)^{(d_1-d_2-d_3)/2}((z-x)^2)^{(d_2-d_1-d_3)/2}\;.
\end{eqnarray*}
The factors $(X^5+X^6)^{-d_1}$, $(Y^5+Y^6)^{-d_2}$, and $(Z^5+Z^6)^{-d_3}$ are canceled by similar factors in the relation (22) between the $\varphi$s and $\Phi$s, leaving us with a three-point function  $\left<\varphi_1(x)\varphi_2(y)\varphi_3(z)\right>_0$ proportional to 
\begin{equation}
((x-y)^2)^{(d_3-d_1-d_2)/2}((y-z)^2)^{(d_1-d_2-d_3)/2}((z-z)^2)^{(d_2-d_1-d_3)/2}\;,
\end{equation}
another known result.

\vspace{10 pt}

\noindent
{\bf B. Vector Fields}

\vspace{10pt}

We next turn to vector fields.  The two-point function of the six-vector fields $V^K_1(X)$ and $V^L_2(Y)$ must be a linear combination of the two  tensors that vanish when contracted with either $X_K$ or $Y_L$:
$$\eta^{KL}-\frac{Y^K X^L}{X\cdot Y}\;,~~~~~~ X^K Y^L\;,$$
with coefficients that are functions only of $X\cdot Y$.
Because $X^K\,e^\mu_K(x)=0$, the  second of these makes no contribution to the four-dimensional Green's function, and can be ignored.  Each term in the first transverse tensor contains zero net factors of $X$ and $Y$, while the scaling condition (8) requires that the two-point function be of order $-d_1$ in $X$ and of order $-d_2$ in $Y$, so we see again that the two-point function vanishes unless $d_1=d_2\equiv d$,  in which case it is proportional to
$$(X\cdot Y)^{-d}\left(\eta^{KL}-\frac{Y^K X^L}{X\cdot Y}\right)\;,$$
 with a constant coefficient.  Using Eq.~(17), we see that the four-dimensional Green's function $\left<v^\mu(x)v^\nu(y)\right>_0$ is proportional to
$$(X^5+X^6)^d(Y^5+Y^6)^d(X\cdot Y)^{-d}e^\mu_K(x)e^\nu_L(Y)\left(\eta^{KL}-\frac{Y^K X^L}{X\cdot Y}\right) $$
Now, we note that 
\begin{equation}
e^\mu_K(x)e^\nu_L(y)\eta^{KL}=\eta^{\mu\nu}\;,
\end{equation}
and
\begin{equation}
Y^K e^\mu_K(x)=Y^\mu-x^\mu (Y^5+Y^6)=(Y^5+Y^6)(y^\mu-x^\mu)
\end{equation}
and likewise $X^K e^\mu_K(y)=(X^5+X^6)(x^\mu-y^\mu)$.   Eq.~(21) then shows that the factors $(X^5+X^6)$ and $(Y^5+Y^6)$ all cancel, leaving us with the result that 
$\left<v^\mu(x)v^\nu(y)\right>_0$  is proportional to
\begin{equation}
((x-y)^2)^{-d}\left(\eta^{\mu\nu}-2\frac{(x-y)^\mu (x-y)^\nu}{(x-y)^2}\right)\;.
\end{equation}
Here the conformal dimensionality $d$ is arbitrary, but if we now impose the further condition that these vectors are conserved currents, we find that $d$ must have the canonical value $d=3$.

\vspace{10 pt}

\noindent
{\bf C. Symmetric Second-Rank Tensor  Fields}

\vspace{10pt}

The two-point function of two symmetric six-tensors $T_I^{KL}(X)$ and $T_2^{MN}(Y)$ is required by $SO(4,2)$ invariance and the transversality condition (16) to  be a linear combination of the transverse tensors
$$
 \left(\eta^{KM}-\frac{Y^KX^M}{X\cdot Y}\right)\left(\eta^{LN}-\frac{Y^LX^N}{X\cdot Y}\right) 
+\left(\eta^{LM}-\frac{Y^LX^M}{X\cdot Y}\right)\left(\eta^{KN}-\frac{Y^KX^N}{X\cdot Y}\right)
$$
$$
\left(\eta^{KL}-\frac{Y^KX^L+Y^LX^K}{X\cdot Y}\right)\left(\eta^{MN}-\frac{X^MY^N+Y^MX^N}{X\cdot Y}\right)
$$
and 
$$ X^KX^LY^MY^N \;,$$
in all three  cases with coefficients that are functions only of the scalar $X\cdot Y$.  
Each term in these three  tensors (including their  coefficients) has equal numbers of factors of $X$ and $Y$, while the scaling condition (15) requires the number of factors of $X$ and $Y$ to equal $-d_1$ and $-d_2$, respectively, so we must have $d_1=d_2\equiv d$, just as for scalars and vectors.  
Because $X^Ke^\mu_K(x)=Y^Me^\mu_M(y)=0$, the third of these tensors makes no contribution to the four-dimensional two-point function, and will therefore be ignored.  So the six-dimensional Green's function must be a linear combination of the first two tensors, with coefficients proportional to $(X\cdot Y)^{-d}$.  
Using Eqs.~(21), (24) and (25), the two-point function of the four-dimensional tensors defined by Eq.~(17) is then
\begin{eqnarray}
&&\left<t^{\mu\nu}(x)t^{\rho\sigma}(y)\right>_0 = A[r^2]^{-d}\Bigg[\eta^{\mu\rho}\eta^{\nu\sigma}+\eta^{\mu\sigma}\eta^{\nu\rho}\nonumber \\&&
~~-2\frac{r^\rho r^\nu\eta^{\mu\sigma}+r^\rho r^\mu\eta^{\nu\sigma}+r^\sigma r^\nu\eta^{\mu\rho}+r^\sigma r^\mu\eta^{\nu\rho}}{r^2}\nonumber\\&&~~~~~
+8\frac{r^\rho r^\sigma r^\mu r^\nu}{(r^2)^2}\Bigg]
+B[r^2]^{-d}\eta^{\mu\nu}\eta^{\rho\sigma}\;,
\end{eqnarray}
where $r\equiv x-y$, and  $A$ and $B$ are constants.

So far, $d$ like $A$ and $B$ is an arbitrary number, but all these constants become tightly constrained if we require that the tensor is conserved.  Operating on Eq.~(27) with $\partial/\partial x^\mu$ gives a quantity proportional to 
$$
(2d-8)(r^\rho\eta^{\sigma\nu}+r^\sigma\eta^{\rho\nu})-(4A+2dB)r^\nu\eta^{\rho\sigma}+A(32-8d)\frac{r^\rho r^\sigma r^\nu}{r^2}\;,
$$
so the conservation condition tells us that $d=4$ and $A=-2B$.  These are just the properties we expect for the energy-momentum tensor in a conformally invariant theory --- its canonical dimension is $d=4$, while the condition $A=-2B$ tells us that the tensor is traceless.

\begin{center}
{\bf IV. SPIN0R FIELDS}
\end{center}

We now consider how to convert information about the Green's functions of spinor fields on the hypercone in six-dimensional projective space into information about the Green's functions of spinors in four-dimensional spacetime.  Let's first recall some well-known facts about spinors in six dimensions.

The Clifford algebra for $SO(4,2)$ has a $2^{6/2}=8$-dimensional irreducible representation:
\begin{equation}
\Gamma^\mu=\left(\begin{array}{cc}0 & i\gamma_5\gamma^\mu \\ i\gamma_5\gamma^\mu & 0\end{array}\right),~~~
\Gamma^5=\left(\begin{array}{cc}0 & \gamma_5 \\ \gamma_5 & 0\end{array}\right),~~~
\Gamma^6=\left(\begin{array}{cc}0 & 1 \\ -1 & 0\end{array}\right)\,,
\end{equation}
which obeys the anticommutation relations
\begin{equation}
\left\{\Gamma^K,\Gamma^L\right\}=2\eta^{KL}\;.
\end{equation}
(Here $\gamma_\mu$ is the usual $4\times 4$ Dirac matrix,\footnote{Our notation for Dirac matrices is the same as used in [11].} and $\gamma_5\equiv -i\gamma^0\gamma^2\gamma^2\gamma^3$.)    From these matrices, we can construct the  8-component Dirac representation of the $SO(4,2)$ Lie algebra
\begin{equation}
{\cal J}^{KL}=-\frac{i}{4}\,\Big[\Gamma^K,\Gamma^L\Big]
\end{equation}
for which
\begin{equation}
i\Big[{\cal J}^{KL},\Gamma^M\Big]=\Gamma^K\eta^{LM}-\Gamma^L\eta^{KM}\;,
\end{equation}
and so
\begin{equation}
i\Big[{\cal J}^{KL},{\cal J}^{MN}\Big]=\eta^{LM}{\cal J}^{KN}-\eta^{KM}{\cal J}^{LN}
-\eta^{LN}{\cal J}^{KM}+\eta^{KN}{\cal J}^{LM}\;.
\end{equation}
Explicitly,
\begin{eqnarray}
 & {\cal J}^{\mu\nu}=\left(\begin{array}{cc} j^{\mu\nu} & 0 \\ 0  & j^{\mu\nu}\end{array}\right),~~~& 
{\cal J}^{5\mu}=\frac{1}{2}\left(\begin{array}{cc} \gamma^\mu & 0 \\ 0 & \gamma^\mu\end{array}\right),\nonumber \\&
{\cal J}^{6\mu}=\frac{1}{2}\left(\begin{array}{cc} \gamma_5\gamma^\mu & 0 \\ 0 & -\gamma_5\gamma^\mu\end{array}\right),~~~ & 
{\cal J}^{56}=\frac{i}{2}\left(\begin{array}{cc} \gamma_5 & 0 \\ 0 & -\gamma_5\end{array}\right)\;,
\end{eqnarray}
where $j^{\mu\nu}$ is the Dirac representation of the Lorentz group Lie algebra:
\begin{equation}
j^{\mu\nu}=-\frac{i}{4}\Big[\gamma^\mu,\gamma^\nu\Big]\;.
\end{equation}
The block diagonal form of the matrices (33) indicates that this representation of the Lie algebra of $SO(4,2)$ is reducible, the top and bottom blocks furnishing the two different irreducible four-component spinor representations 
of the Lie algebra of $SO(4,2)$.  

The 8-component spinor fields in six dimensions have an $SO(4,2)$ transformation given by the commutation relations
\begin{equation}
i[J^{KL},\Psi^n(X)]=\left(X^L\frac{\partial}{\partial X_K}-X^K\frac{\partial}{\partial X_L}\right)\Psi^n(X)-i\left({\cal J}^{KL}\right)^n{}_m\Psi^m(X)\;.
\end{equation}
We note that the matrices $\Gamma^K$ and ${\cal J}^{KL}$ obey  reality conditions
\begin{equation}
\Big(\Gamma^K\Big)^\dagger=-b\Gamma^K b\;,~~~\Big({\cal J}^{KL}\Big)^\dagger=b{\cal J}^{KL}b\;,~~~b\equiv \left(\begin{array}{cc} \gamma^0\gamma_5 & 0 \\ 0 & \gamma^0\gamma_5\end{array}\right)=b^{-1}\;,
\end{equation}
so the adjoint of Eq.~(35) gives
\begin{equation}
i[J^{KL},\overline{\Psi}(X)]=\left(X^L\frac{\partial}{\partial X_K}-X^K\frac{\partial}{\partial X_L}\right)\overline{\Psi}(X)+i\overline{\Psi}(X)\,{\cal J}^{KL}\;,
\end{equation}
where
\begin{equation}
\overline{\Psi}(X)\equiv \Psi^\dagger(X)\, b\;.
\end{equation}
We can therefore form six-tensors from bilinears in $\Psi$: For any $8\times 8$ matrix $M$, we have 
\begin{eqnarray}
&&i\left[J^{KL},\Big(\overline{\Psi}(X)M\Psi(X)\Big)\right]=\left(X^L\frac{\partial}{\partial X_K}-X^K\frac{\partial}{\partial X_L}\right)\Big(\overline{\Psi}(X)M\Psi(X)\Big)\nonumber\\&&~~~~~~~~+i\Big(\overline{\Psi}(X)\,[{\cal J}^{KL},M]\Psi(X)\Big)\;,
\end{eqnarray}
so for instance $\Big(\overline{\Psi}(X)\Gamma^K\Psi(X)\Big)$ is a vector field, $\Big(\overline{\Psi}(X){\cal J}^{KL}\Psi(X)\Big)$ is an antisymmetric tensor, etc.

As in the case of tensor fields, we assume that $\Psi(X)$ obeys a scaling law,
\begin{equation}
\Psi(\lambda X) =\lambda^{-d+1/2}\Psi(X)\;
\end{equation}
so that $(X^5+X^6)^{d-1/2}\Psi(X)$ is a function only of ratios of the $X^K$.  So far, $d-1/2$ is just some unknown number; the reason for writing it in this form will become apparent soon.  With $X^5-X^6$ eliminated in favor of $X^5+X^6$ and $X^\mu X_\mu$ by use of Eq.~(6), we can regard $(X^5+X^6)^{d-1/2}\Psi(X)$ as a function only of the coordinate $x^\mu$ given by Eq.~(8):
\begin{equation}
(X^5+X^6)^{d-1/2}\Psi(X)\equiv\zeta(x)\;
\end{equation}
It will be convenient to separate $\Psi(x)$ and $\zeta(x)$ into four-component segments
\begin{equation}
\Psi(x)=\left(\begin{array}{c} \Psi_+(x) \\ \Psi_-(x)\end{array}\right)\;,~~~
\zeta_\pm(x)=(X^5+X^6)^{d-1/2}\Psi_\pm(X)\;.
\end{equation}
Eq.~(33) shows that the $\Psi_\pm$ transform according to the two fundamental spinor irreducible representations of $SO(4,2)$.
Although the $\zeta_\pm(x)$ are functions only of $x^\mu$, neither of these four-component fields  have the right conformal (or even translation) transformation properties (1)--(4) to serve as conventional four-dimensional spinor fields, but they will be ingredients in the construction of such fields.

Using Eqs.~(35) and (33), we can work out the commutators of the $\zeta_\pm$  fields with the generators $J^{\mu\nu}$ of Lorentz transformations; the generators $P^\mu=J^{5\mu}+J^{6\mu}$ of translations, the generators $K^\mu=J^{6\mu}-J^{5\mu}$ of special conformal transformations, and the generator $S=-J^{56}$ of scale transformations:
\begin{eqnarray}
&&i[J^{\mu\nu},\zeta_\pm(x)]=\left(x^\nu\frac{\partial }{\partial x_\mu}-x^\mu\frac{\partial }{\partial x_\nu}\right)\zeta_\pm(x) -i j^{\mu\nu}\zeta_\pm(x)\;,\\
&& i[P^\mu,\zeta_\pm(x)]=-\frac{\partial}{\partial x_\mu}\zeta_\pm(x)-\frac{i}{2}(1\pm \gamma_5)\gamma^\mu \zeta_\pm(x)\;,\\ 
&& i[K^\mu,\zeta_\pm(x)]=\left(2x^\mu x^\lambda\frac{\partial}{\partial x^\lambda}-x^2\frac{\partial}{\partial x_\mu}+(2d-1) x^\mu\right)\zeta_\pm(x) \nonumber\\&&~~~+\frac{i}{2}(1\mp \gamma_5)\gamma^\mu\zeta_\pm(x)\;,\\
&& i[S,\zeta_\pm(x)]=\left(x^\lambda \frac{\partial}{\partial x^\lambda}+d-\frac{1}{2}\right)\zeta_\pm(x)\mp \frac{1}{2}\gamma_5\zeta_\pm(x)\;.
\end{eqnarray}
The second terms in Eqs. (44) through (46) are very different from the matrix terms in the commutation relations (1)--(4) of general fields in four-dimensions.  In particular, the presence of a matrix term in the commutation relation (44) shows that $\zeta_\pm(x)$ does not have the usual transformation rule under translations.  In order to construct suitable four-dimensional spinor fields, we must impose a condition on  $\Psi(X)$ analogous to the transversality condition imposed on tensors in Section II, and we must apply a projection matrix to $\zeta_\pm(X)$, analogous to the quantities $e^\mu_K(x)$ in Eq.~(17).

First, to eliminate the matrix term in Eq.~(44), we define a pair of chiral fields
\begin{equation}
\psi_\pm(x)\equiv \frac{1}{2}(1\mp \gamma_5)\,\zeta_\pm(x)\;.
\end{equation}
Because $\gamma_5$ commutes with $j^{\mu\nu}$, multiplying Eq.~(43) with $(1\mp\gamma_5)/2$ gives the same Lorentz transformation rule:
\begin{equation}
i[J^{\mu\nu},\psi_\pm(x)]=\left(x^\nu\frac{\partial }{\partial x_\mu}-x^\mu\frac{\partial }{\partial x_\nu}\right)\psi_\pm(x) -i j^{\mu\nu}\psi_\pm(x)\;,
\end{equation}
while multiplying Eq.~(44) with $(1\mp\gamma_5)/2$ gives what is now a conventional transformation under spacetime translations:
\begin{equation}
i[P^\mu,\psi_\pm(x)]=-\frac{\partial}{\partial x_\mu}\psi_\pm(x)\;.
\end{equation}
When we multiply Eq.~(46) with $(1\mp\gamma_5)/2$, the second term becomes just $\psi_\pm(x)/2$, canceling the $-1/2$ in the first term:
\begin{equation}
i[S,\psi_\pm(x)]=\left(x^\lambda \frac{\partial}{\partial x^\lambda}+d\right)\,\psi_\pm(x)\;.
\end{equation}
This is why we wrote the scaling relation for fermions in the form (40); Eq.~(50) shows that with this form of the scaling relation, $d$ is the conformal dimension of the spinor fields.  Finally, multiplying the commutation relation (45) with $(1\mp\gamma_5)/2$ gives 
\begin{eqnarray}
&&i[K^\mu,\psi_\pm(x)]=\left(2x^\mu x^\lambda\frac{\partial}{\partial x^\lambda}-x^2\frac{\partial}{\partial x_\mu}+(2d-1) x^\mu\right)\psi_\pm(x) \nonumber\\&&~~~+i\gamma^\mu\chi_\pm(x)\;,
\end{eqnarray}
where $\chi_\pm$ is the opposite-chirality part of $\zeta_\pm$:
\begin{equation}
\chi_\pm(x)\equiv \frac{1}{2}(1\pm \gamma_5)\,\zeta_\pm(x)\;.
\end{equation}
This is still very different from the desired transformation rule under special conformal transformations.

To proceed, we must impose a transversality condition on the spinor fields $\Psi(X)$ in six dimensions.  The natural such condition is
\begin{equation}
X_K\Gamma^K\Psi(X)=0\;.
\end{equation}
This manifestly respects $SO(4,2)$ invariance, and it is consistent with the fact that $(X\cdot\Gamma)^2=(X\cdot X)=0$, so that zero is the sole eigenvalue of $X\cdot\Gamma$.  Eq.~(53)  has the immediate consequence that the vector field $(\overline{\Psi}\Gamma^K\Psi)$ obeys the same transversality condition $X_K(\overline{\Psi}\Gamma^K\Psi)=0$ that we imposed on vector fields in Section II.  The same transversality holds for the other vector field $(\overline{\Psi}\Gamma_7\Gamma^K\Psi)$, where
\begin{equation}
 \Gamma_7\equiv -i\Gamma^0\Gamma^1\Gamma^2\Gamma^3\Gamma^5\Gamma^6=\left(\begin{array}{cc} 1 & 0 \\ 0 & -1\end{array}\right)\;,
\end{equation}
and also for the antisymmetric tensors $(\overline{\Psi}[\Gamma^K, \Gamma^L]\Psi)$ and $(\overline{\Psi}\Gamma_7[\Gamma^K, \Gamma^L]\Psi)$.
The only other six-dimensional tensors that can be formed from bilinears in $\Psi(X)$ are the totally antisymmetric tensors of third rank
$$ (\overline{\Psi}\Gamma^{[K} \Gamma^L \Gamma^{M]}\Psi)\;,~~~~(\overline{\Psi}\Gamma_7\Gamma^{[K} \Gamma^L \Gamma^{M]}\Psi)\;,$$
the square brackets indicating antisymmetrization.  These are not strictly transverse; instead, Eq.~(53) gives
$$ X_K(\overline{\Psi}\Gamma^{[K} \Gamma^L \Gamma^{M]}\Psi)=X^L(\overline{\Psi}\Gamma^M\Psi)-X^M(\overline{\Psi}\Gamma^L\Psi)\;,
$$
and similarly  for $X_K(\overline{\Psi}\Gamma_7\Gamma^{[K} \Gamma^L \Gamma^{M]}\Psi)$.  If we think of these tensors as three-forms 
$$(\overline{\Psi}\Gamma^{[K} \Gamma^L \Gamma^{M]}\Psi)dX_K\,dX_L\,dX_M\;,~~~~  (\overline{\Psi}\Gamma_7\Gamma^{[K} \Gamma^L \Gamma^{M]}\Psi)dX_K\,dX_L\,dX_M$$
 with anticommuting differentials $dX^K$ tangent to the hypercone (6), so that $X^K dX_K=0$, then these 3-forms are transverse, in the sense that they vanish if we replace any $dX^K$ with $X^K$.  But the real justification for the transversality condition (53) is that, as we shall now see, it gives the results we need in four dimensions.

By multiplying the transversality condition Eq.~(53) with the matrix 
$$ \left(\begin{array}{cc} 1-\gamma_5 & 0 \\ 0 & 1+\gamma_5\end{array}\right)$$
we find a simple formula for $\chi_\pm$ in terms of $\psi_\pm$:
\begin{equation}
\chi_\pm=-ix_\nu\gamma^\nu\psi_\pm\;.
\end{equation}
Thus the last term in Eq.~(51) is
$$i\gamma^\mu \chi_\pm=\gamma^\mu\gamma^\nu x_\nu\psi_\pm=\Big(x^\mu+2ij^{\mu\nu} x_\nu\Big)\psi_\pm\;.$$
The special conformal transformation rule (51) thus reads
\begin{eqnarray}
&&i[K^\mu,\psi_\pm(x)]=\left(2x^\mu x^\lambda\frac{\partial}{\partial x^\lambda}-x^2\frac{\partial}{\partial x_\mu}+2d x^\mu\right)\psi_\pm(x) \nonumber\\&&~~~+2i j^{\mu\nu}x_\nu\psi_\pm(x)\;.
\end{eqnarray}
Eqs.~(48)--(50) and (56)  show that the fields $\psi_\pm(x)$ are conventional four-dimensional Dirac fields, satisfying the commutation relations (1)--(4) with the generators of the conformal group, and with conformal dimension  $d$.  The other fields $\chi_\pm$ have no obvious physical interpretation.  Of course, we can assemble the chiral fields $\psi_\pm$ into a four-component Dirac field
\begin{equation}
\psi(x)=\psi_+(x)+\psi_-(x)=(X^5+X^6)^{d-1/2}\left[\left(\frac{1-\gamma_5}{2}\right)\Psi_+(X)+
\left(\frac{1+\gamma_5}{2}\right)\Psi_-(X)\right]\;.
\end{equation}
It is this form of the spinor field that will be used to work out the consequences of conformal symmetry for Green's functions involving spinor fields.

By combining the methods of this section and of Section II, we can see that a field $\Psi^{K_1\cdots K_r}(X)$ with tensor indices as well as an 8-component spinor index, if subjected to the transversality conditions,
$$
X_{K_1}\Psi^{K_1\cdots K_r}(X)=\dots=X_{K_r}\Psi^{K_1\cdots K_r}(X)=(X\cdot \Gamma)\Psi^{K_1\cdots K_r}(X)=0
$$
yields a spinor-tensor in four dimensions
\begin{eqnarray*}
&& \psi^{\mu_1\cdots \mu_r}(x)=(X^5+X^6)^{d-1/2}e^{\mu_1}_{K_1}(x)\cdots e^{\mu_r}_{K_r}(x) \nonumber\\&&\times \left[\frac{(1-\gamma_5)}{2}\Psi_+^{K_1\cdots K_r}(X)+\frac{(1+\gamma_5)}{2}\Psi_-^{K_1\cdots K_r}(X)\right]\;,
\end{eqnarray*}
(where $\Psi_+$ and $\Psi_-$ are the upper and lower four components of $\Psi$, with $\Gamma_7=+1$ and $\Gamma_7=-1$, respectively), which transforms under conformal transformations according to Eqs.~(1)--(4), with conformal dimensionality $d$.

\vspace*{30pt}

\begin{center}
{\bf V. SPINOR APPLICATIONS}
\end{center}

First let's consider the Green's function $\left<\psi_1(x)\,\overline{\psi}_2(y)\right>$, where $\overline{\psi}\equiv \psi^\dagger \gamma^0\gamma_5$.  Invariance under $SO(4,2)$ tells us that the corresponding two point function of $\Psi_1(X)$ and $\overline{\Psi}_2(Y)$ in six dimensions must be a linear combination
$$A+B(X\cdot\Gamma)+C(Y\cdot \Gamma) +D[X\cdot \Gamma,Y\cdot \Gamma] \;,$$
with $A$, $B$, $C$, and $D$ all functions only of the scalar $X\cdot Y$.  (Here we are ignoring the possibility of including terms involving the matrix  $\Gamma_7$.  We will consider such terms presently.)  The transversality condition that $(X\cdot\Gamma)\Psi_1(X)=0$ tells us that  $C=0$ and $A=2D\,X\cdot Y$, while the condition that $\overline{\Psi}_2(Y)(Y\cdot \Gamma)=0$
tells us $B=0$ and, again, $A=2D\,X\cdot Y$.  So the six-dimensional Green's function must have the form
$$A\,\left(1+\frac{[X\cdot \Gamma,Y\cdot \Gamma]}{2X\cdot Y}\right)=\frac{A\,(X\cdot\Gamma)\,(Y\cdot \Gamma)}{(X\cdot Y)}\;.$$
Every term here has equal numbers of factors of $X^K$ and $Y^K$ (including those in $A$), while the scaling condition (40) tells us that the Green's function must be of order $-d_1+1/2$ in $X^K$ and of the  order $-d_2+1/2$ in $Y^K$, so we must have $d_1=d_2\equiv d$, and the whole Green's function must be proportional to
\begin{equation}
(X\cdot Y)^{1/2-d}\left(1+\frac{[X\cdot \Gamma,Y\cdot \Gamma]}{2X\cdot Y}\right)\;,
\end{equation}
with a constant proportionality coefficient.

From Eqs.~(30) and (33), we find
$$
[X\cdot \Gamma,Y\cdot \Gamma]=4iX_KY_L{\cal J}^{KL}=4i\left(\begin{array}{cc} M_+ & 0 \\ 0 & M_-\end{array}\right)\;,
$$
where 
\begin{eqnarray*} M_\pm &=& j^{\mu\nu}X_\mu Y_\nu +\frac{1}{2}(1\pm\gamma_5)\gamma^\mu (X_5 Y_\mu-Y_5 X_\mu)\\&&\pm
\gamma_5\gamma^\mu(X_6Y_\mu-Y_6X_\mu)\pm \frac{i}{2}\gamma_5 (X_5Y_6-Y_5X_6) \;.
\end{eqnarray*}
From Eq.~(57), we then have
\begin{eqnarray*}
&&\left<\psi_1(x)\overline{\psi}_2(y)\right>\propto (X^5+X^6)^{d-1/2}(Y^5+Y^6)^{d-1/2}(X\cdot Y)^{-d-1/2}\\ &&~~~\times \sum_\pm\left(\frac{1\mp\gamma_5}{2}\right)M_\pm\left(\frac{1\pm\gamma_5}{2}\right)\;;.
\end{eqnarray*}
Only the vector and axial vector terms in $M_\pm$ survive, so this simplifies to
\begin{eqnarray*}
&&\left<\psi_1(x)\overline{\psi}_2(y)\right>\propto (X^5+X^6)^{d-1/2}(Y^5+Y^6)^{d-1/2}(X\cdot Y)^{-d-1/2}\\ &&~~~\times 
\gamma^\mu\Big((X^5+X^6)Y_\mu-(Y^5+Y^6)X_\mu\Big)\;.
\end{eqnarray*}
From (8) and (21), we have then
\begin{equation}
\left<\psi_1(x)\overline{\psi}_2(y)\right>\propto \Big((x-y)^2\Big)^{-d-1/2}\gamma^\mu (x_\mu-y_\mu)\;.
\end{equation}
This is of course just what we should expect in a Poincar\'{e} invariant and scale invariant theory with spinor fields of equal dimensionality $d$.

Now let us return to the possibility of including the matrix $\Gamma_7$ defined by Eq.~(54) in the six-dimensional
Green's function.  That is, we consider the possibility of multiplying Eq.~(58) with a factor $(1+\alpha\Gamma_7)$, with some arbitrary $\alpha$, so that the Green's function in six dimensions is proportional to
\begin{equation}
(1+\alpha\Gamma_7)(X\cdot Y)^{1/2-d}\left(1+\frac{[X\cdot \Gamma,Y\cdot \Gamma]}{2X\cdot Y}\right)\;.
\end{equation}
The effect is to multiply 
the terms $M_\pm$  with $(1\pm \alpha)$, so that the Green's function (59) becomes
\begin{equation}
\left<\psi_1(x)\overline{\psi}_2(y)\right>\propto \Big((x-y)^2\Big)^{-d-1/2}(1-\alpha\gamma_5)\gamma^\mu (x_\mu-y_\mu)\;.
\end{equation}
This is allowed by $SO(4,2)$ invariance, since $\Gamma_7$ commutes with all the generators ${\cal J}^{KL}$, but it is not allowed in a theory that is invariant under $O(4,2)$, since $\Gamma_7$ changes sign under transformations (9) with ${\rm Det}\Lambda=-1$.  In particular, $\Gamma_7$ terms seem to be ruled out if we impose invariance under the inversion $x^\mu\mapsto -x^\mu/x^2$, which just amounts to the reflection that changes the sign of $X^6$ and leaves all other $X^K$ unchanged. 
 
The presence of a $\Gamma_7$ term in the six-dimensional Green's function (60) or a $\gamma_5$ term in the corresponding four-dimensional Green's function (61) does not in itself violate invariance under $O(4,2)$, because we can eliminate these terms by a redefinition of the fermion fields.  It is only necessary to replace $\Psi$ with 
\begin{equation}
\Psi'=\left[(1+\alpha)^{-1/2}\left(\frac{1+\Gamma_7}{2}\right)+(1-\alpha)^{-1/2}\left(\frac{1-\Gamma_7}{2}\right)\right]\Psi
\end{equation}
so that instead of Eq.~(56) we have
\begin{equation}
\psi(x)=(X^5+X^6)^{d-1/2}\left[(1+\alpha)^{-1/2}\left(\frac{1-\gamma_5}{2}\right)\Psi_+(X)+
(1-\alpha)^{-1/2}\left(\frac{1+\gamma_5}{2}\right)\Psi_-(X)\right]\;.
\end{equation}
  The real sign of a breakdown of $O(4,2)$ to $SO(4,2)$ is the presence, in one or more Green's functions, of $O(4,2)$-breaking $\Gamma_7$  terms that cannot all be eliminated by redefinition of the fermion fields.

Here is an example.  Consider the Green's function $\left<\psi_1(x)\overline{\psi}_2(y)\varphi(z)\right>_0$ of two fermion and one scalar field, of dimensionality $d_1$, $d_2$, and $d_3$, respectively.  Invariance under $O(4,2)$ would require the corresponding six-dimensional Green's function to take the form
\begin{eqnarray*}
&& A+B(X\cdot\Gamma)+C(Y\cdot\Gamma)+D(Z\cdot\Gamma)+E[X\cdot \Gamma,Y\cdot \Gamma]\\&&~~~~~~~
F[Y\cdot \Gamma,Z\cdot \Gamma]+G[Z\cdot \Gamma, X\cdot \Gamma]+H(X\cdot \Gamma)\,(Z\cdot\Gamma)\,(Y\cdot\Gamma)\;,
\end{eqnarray*}
with $A$, $B$, etc. functions of the scalars $X\cdot Y$, $Y\cdot Z$, and $Z\cdot X$.  (Any other ordering of the $\Gamma$-matrices in the last term would differ only by terms of the same form as those already included.)
This must vanish when we multiply with $X\cdot \Gamma$ on the left; the vanishing of the terms proportional to $[X\cdot \Gamma, Y\cdot \Gamma]$, $[X\cdot \Gamma, Z\cdot \Gamma]$, and $X_KY_LZ_M\,\Gamma^{[K}\Gamma^L\Gamma^{M]}$ gives $C=0$, $D=0$, and $F=0$, while the vanishing of the terms proportional to $X\cdot \Gamma$ gives $A=2EX\cdot Y$.  It must also vanish when we multiply on the right with    $Y\cdot \Gamma$; the vanishing of the terms proportional to $[X\cdot \Gamma, Y\cdot \Gamma]$, $[Y\cdot \Gamma, Z\cdot \Gamma]$, and $X_KY_LZ_M\,\Gamma^{[K}\Gamma^L\Gamma^{M]}$ gives $B=0$, $D=0$, and $G=0$, while the vanishing of the terms proportional to $Y\cdot \Gamma$ again gives $A=2EX\cdot Y$. In both cases the vanishing of terms proportional to the unit matrix gives nothing new.  So we conclude that the Green's function in six dimensions is of the form
$$A\left(1+\frac{[X\cdot \Gamma, Y\cdot \Gamma]}{2X\cdot Y}\right)+H(X\cdot \Gamma)\,(Z\cdot\Gamma)\,(Y\cdot\Gamma)\;.$$
Now, according to the scaling properties of the fields, the total number of factors of $X$, $Y$, and $Z$ must be respectively $-d_1+1/2$, $-d_2+1/2$, $-d_3$, so  
$$A\propto(X\cdot Y)^{-a}(Y\cdot Z)^{-b}(Z\cdot X)^{-c}\;,$$
$$H\propto (X\cdot Y)^{-a-1/2}(Y\cdot Z)^{-b-1/2}(Z\cdot X)^{-c-1/2}\;,$$
where $a+c=d_1-1/2$, $a+b=d_2-1/2$, $b+c=d_3$.    The Green's function for two spinors and a scalar in six dimensions thus takes the form
\begin{eqnarray}
&& (X\cdot Y)^{(d_3-d_1-d_2+1)/2}(Y\cdot Z)^{(d_1-d_2-d_3)/2}(Z\cdot X)^{(d_2-d_3-d_1)/2}\nonumber\\&&~~~\times
\left[a\left(1+\frac{[X\cdot \Gamma, Y\cdot \Gamma]}{2X\cdot Y}\right)+h\frac{(X\cdot \Gamma)\,(Z\cdot\Gamma)\,(Y\cdot\Gamma)}{\sqrt{(X\cdot Y)\,(Y\cdot Z)\,(Z\cdot X)}}\right]\;,
\end{eqnarray}
where $a$ and $h$ are constants.

The contribution of the second term to the four-dimensional Green's function is complicated, and is not needed for the point I wish to make, so I will take $h=0$ in what follows.  Then, following the same arguments as for the two-spinor Green's function, we have
\begin{eqnarray}
&&\left<\psi_1(x)\overline{\psi}_2(y)\varphi(z)\right>_0\propto 
((x-y)^2)^{(d_3-d_1-d_2-1)/2}((y-z)^2)^{(d_1-d_2-d_3)/2}\nonumber\\&&~~~~\times ((z-x)^2)^{(d_2-d_3-d_1)/2}\gamma^\mu (x-y)_\mu\;.
\end{eqnarray}
But in a theory that is invariant under $SO(4,2)$ but not $O(4,2)$, we are free to include a factor $1+\beta\Gamma_7$ multiplying the first term in Eq.~(64), so that (for $h=0$) in place of Eq.~(65) we have
\begin{eqnarray}
&&\left<\psi_1(x)\overline{\psi}_2(y)\varphi(z)\right>_0\propto 
((x-y)^2)^{(d_3-d_1-d_2-1)/2}((y-z)^2)^{(d_1-d_2-d_3)/2}\nonumber\\&&~~~~\times ((z-x)^2)^{(d_2-d_3-d_1)/2}(1-\beta\gamma_5)\gamma^\mu (x-y)_\mu\;.
\end{eqnarray}
Now, by redefining the fermion fields we can eliminate the $1+\alpha\Gamma_7$ factor in the two point function, which eliminates the $\gamma_5$ term in Eq.~(61),  or we can eliminate the $1+\beta\Gamma_7$ factor  in the three-point function, which eliminates the $\gamma_5$ term in Eq.~(66), but unless $\beta=\alpha$ we cannot do both.  We see then that it makes a difference whether we assume invariance under $O(4,2)$, which includes the inversion $x^\mu\mapsto -x^\mu/x^2$, or only invariance under $SO(4,2)$, which does not include the inversion.

\begin{center}
{\bf VI. AdS/CFT}
\end{center}

In the preceeding sections the six-tensors $T^{K_1\cdots K_r}(X)$ and eight-component spinors $\Psi(X)$ were fictions, merely means to the end of calculating Green's functions for fields in four spacetime dimensions.  But $T^{K_1\cdots K_r}(X)$ and  $\Psi(X)$ may also be regarded as actual fields on five-dimensional anti-de Sitter space (AdS$_5$).  This space is the surface of the hypersphere in six dimensions
\begin{equation}
\eta_{KL}X^K X^L=R^2
\end{equation}
with the same metric $\eta_{KL}$ as in Sections I through V, and arbitrary $R> 0$.  It is manifestly maximally symmetric, with isometry group $SO(4,2)$  consisting of the transformations (9).  Tensors $T^{K_1\cdots K_r}(X)$ on AdS$_5$ transform as in Eq.~(12), and without upsetting the isometry can be subject to the transversality condition (16).  We can also introduce 8-component spinor fields $\Psi(X)$ on AdS$_5$, with the same $SO(4,2)$ transformation properties as in Section IV, but we cannot here adopt the transversality condition (53), which requires that $(X\cdot \Gamma)\,\Psi=0$, because on the hypersphere we have
$$ (X\cdot \Gamma)^2=X\cdot X=R^2\;,$$
and so the only eigenvalues of $X\cdot \Gamma$ are $R$ and $-R$.
  But we can instead adopt the $SO(4,2)$-invariant condition
\begin{equation}
X\cdot\Gamma\, \Psi(X)=R\Psi(X)\;.
\end{equation}
There is no loss of generality in taking the coefficient of $\Psi(X)$ on the right-hand side to  be $R$ rather than $-R$, because if $\Psi(X)$ satisfies Eq.~(68), then $\Gamma_7\Psi(X)$ satisfies the same constraint with $R$ replaced with $-R$.

Of course, $X^K$ and $\lambda X^K$ here can not both be on the hypersphere (67) except for $\lambda=\pm 1$, so we can not impose a scale invariance condition like (15) here.  But in the limit that some components $X^K$ become much larger than $R$,  with the ratios of all components held fixed, the hypersphere (67) effectively becomes the hypercone (6), and the constraint (68) on spinor fields effectively becomes the transversality condition (53).  The AdS/CFT conjecture deals with fields on AdS$_5$ that approach  
c-number values $T_\infty^{K_1\cdots K_r}(X)$ or $\Psi_\infty(X)$ in this limit, satisfying  scaling conditions of the form
\begin{equation}
T_\infty^{K_1\cdots K_r}(\lambda X)=\lambda^{a} T_\infty^{K_1\cdots K_r}(X)\;,~~~
\Psi_\infty(\lambda X)=\lambda^{a}\Psi_\infty(X)\;.
\end{equation}
Of particular interest are massless degrees of freedom, represented by fields with $a=0$; massive degrees of freedom generally have $a<0$.

We know from the work of Sections II and IV that, from such asymptotic fields 
$T_\infty^{K_1\cdots K_r}(X)$ and $\Psi_\infty(X)$, we can form tensor fields (17) and spinor fields (57) in four dimensions that transform as usual under the four-dimensional conformal group, with conformal dimensions $d=-a$ for tensors of any rank and $d-1/2=-a$ for spinors, or spinor-tensors of any rank.  In particular, in the important case $a=0$ for which fields approach finite limits on the boundary $X\rightarrow \infty $ of AdS$_5$, as well known a tensor current on the boundary must have conformal dimension $d=0$, and the four-dimensional tensor field with which it interacts must therefore have dimensionality $d=4$, the expected dimensionality for the energy-momentum tensor in conformally-invariant theories.   On the other hand, a spinor or spinor-tensor field, which arises from a spinor or spinor-tensor field on AdS$_5$ that approaches a finite value on the boundary, has  $d=1/2$, so the  four-dimensional  spinor or spinor-tensor fields with which these fields interact  must then have dimensionality $7/2$,  the correct expected dimensionality for the supersymmetry current in conformally invariant supersymmetric theories.

\begin{center}
{\bf ACKNOWLEDGMENTS}
\end{center}

I am grateful for discussions with J. Distler and J. Meyers, and for correspondence  with I. Bars, A. Chodos, H. Kastrup, T. Okuda, S. Ferrara, and A. Waldron.  This material is based upon work supported by the National Science Foundation under Grant No. PHY-0455649 and with support from The Robert A. Welch Foundation, Grant No. F-0014.

\renewcommand{\theequation}{A.\arabic{equation}}
\setcounter{equation}{0}

\begin{center}
{\bf APPENDIX}
\end{center}

This Appendix will justify the claim made in Section II, that the usual conformal transformation rules of tensor fields just amount to the statement that  under general conformal transformations a tensor of rank $r$ and conformal dimensionality $d$ transforms as a tensor density of weight given by Eq.~(17):
\begin{equation}
t^{\mu_1\mu_2\cdots\mu_r}(x)\mapsto \left|\frac{\partial x}{\partial x'}\right|^{-(r+d)/4}
\frac{\partial x^{\mu_1}}{\partial x'^{\nu_1}}\frac{\partial x^{\mu_2}}{\partial x'^{\nu_2}}\cdots \frac{\partial x^{\mu_r}}{\partial x'^{\nu_r}}t^{\nu_1\nu_2\cdots\nu_r}(x')\;,
\end{equation}
 where $|\partial x/\partial x'|$ is the determinant of the matrix $\partial x^\mu/\partial x'^\nu$.  This is trivial for Lorentz transformations and translations.  For the  scale transformation $x'^\mu=(1+b)x^\mu$, Eq.~(A.1) gives
\begin{equation}
t^{\mu_1\mu_2\cdots\mu_N}(x)\mapsto (1+b)^{d}t^{\mu_1\mu_2\cdots\mu_N}\Big((1+b)x\Big)
\end{equation}
which for infinitesimal $b$ is the same as the scale transformation rule (4).
Similarly, for an infinitesimal  special conformal transformation 
$$ x^\mu\mapsto x'^\mu=x^\mu +2(x\cdot c)x^\mu-c^\mu x^2 \;,$$
we have
$$\frac{\partial x^\mu}{\partial x'^\nu}=\delta^\mu_\nu-2(x\cdot c)\delta^\mu_\nu-2\Big(x^\mu c_\nu-c^\mu x_\nu\Big)\;,~~~
\left|\frac{\partial x}{\partial x'}\right|=1-8(x\cdot c)
$$
so here Eq.~(A.1) reads
\begin{eqnarray}
&&t^{\mu_1\mu_2\cdots\mu_r}(x)\mapsto t^{\mu_1\mu_2\cdots\mu_r}(x)+2d(x\cdot c)t^{\mu_1\mu_2\cdots\mu_r}(x)\nonumber\\&&-(2x^{\mu_1}c_\nu-2c^{\mu_1}x_\nu)t^{\nu\mu_2\cdots\mu_r}(x)
+\cdots-(2x^{\mu_r}c_\nu-2c^{\mu_r}x_\nu)t^{\mu_1\mu_2\cdots\nu}(x)\nonumber\\&&+(2(x\cdot c)x^\mu-c^\mu x^2)\partial_\mu t^{\mu_1\mu_2\cdots\mu_r}(x)\;.
\end{eqnarray}
This is the same as the transformation rule (3) (contracted with $c^\nu$), with Lorentz transformation matrix $j^{\rho\sigma}$ given by Eq.~(19).

\begin{center}
{\bf REFERENCES}
\end{center}

\begin{enumerate}
\item See, e.g., E. J. Schreier, Phys. Rev. D 3, 980 (1971).  For a review, see E. S. Fradkin and M. Ya. Nalchik, Phys. Rept. 44, 249 (1978).
\item P. A. M. Dirac, Ann. Math. 37, 429 (1936).
\item G. Mack and A. Salam, Ann. Phys. (New York) 53, 174 (1969).
\item H. A. Kastrup, Ann. Phys. (Berlin) 17, 631 (2008).
\item I. Bars, Phys. Rev. D 62, 046007 (2000); Phys. Rev. D64, 045004 (2001); Phys. Rev. D 74, 085019 (2006); Phys. Rev D77, 125027 (2008); Phys. Rev. D79, 085021 (2009).
\item S. Ferrara, A. F. Grillo, and R. Gatto, Ann. Phys. (New York) 76, 161 (1973).
\item S. Ferrara, Nucl. Phys. B77, 413 (1974).
\item See for instance L. Cornalba, M. S. Costa, J. Penedones, and R. Schiappa, J. High Energy Phys. 0708, 019 (2007) (which however considers only scalar fields).  
\item J. Maldacena, Adv. Theor. Math. Phys. 2, 231 (1998).
\item E. Witten, Adv. Theor. Math. Phys. 2, 253 (1998).
\item S. Weinberg, {\it The Quantum Theory of Fields} (Cambridge University Press, Cambridge, 1995), Sec. 5.4.
\end{enumerate}
\end{document}